\documentclass[lettersize,journal]{IEEEtran}
\usepackage[T1]{fontenc}
\usepackage{cite}
\usepackage{url}
\usepackage{amsmath,amssymb,amsfonts}
\usepackage{graphicx}
\usepackage{textcomp}
\usepackage{xcolor}
\usepackage{hyperref}
\usepackage{bm}
\usepackage{amsthm}
\usepackage{enumerate}
\usepackage{booktabs}
\usepackage{diagbox}
\usepackage{supertabular}
\usepackage{subfigure}
\usepackage{caption}
\usepackage{algorithmicx}
\usepackage[linesnumbered,ruled,vlined]{algorithm2e}

\begin{document}
	
\pagenumbering{arabic}

\title{Resource-efficient Generative Mobile Edge Networks in 6G Era: Fundamentals, Framework and Case Study}
\author{Bingkun Lai, Jinbo Wen, Jiawen Kang*, Hongyang Du, Jiangtian Nie,  Changyan Yi, \\ Dong In Kim, \textit{IEEE Fellow}, Shengli Xie, \textit{IEEE Fellow}
\thanks{B. Lai, J. Kang, and S. Xie are with the School of Automation, Guangdong University of Technology, Guangzhou 510006, China, and also with the Guangdong-Hong Kong-Macao Joint Laboratory for Smart Discrete Manufacturing, Guangzhou 510006, China (e-mail: bingkunlai@163.com; kavinkang@gdut.edu.cn; shlxie@gdut.edu.cn). J. Wen and C. Yi are with the College of Computer Science and Technology, Nanjing University of Aeronautics and Astronautics, China (e-mail: jinbo1608@163.com; changyan.yi@nuaa.edu.cn). H. Du and J. Nie are with the School of Computer Science and Engineering, Nanyang Technological University, Singapore (e-mail: hongyang001@e.ntu.edu.sg;jnie001@e.ntu.edu.sg). 
 D. I. Kim is with the Department of Electrical and Computer Engineering, Sungkyunkwan University, Suwon 16419, South Korea (e-mail: dikim@skku.ac.kr).

\textit{*Corresponding author: Jiawen Kang.}
 }
 }
	


	\maketitle
	\pagestyle{headings}

	\begin{abstract}
         As the next-generation wireless communication system, Sixth-Generation (6G) technologies are emerging, enabling various mobile edge networks that can revolutionize wireless communication and connectivity. By integrating Generative Artificial Intelligence (GAI) with mobile edge networks, generative mobile edge networks possess immense potential to enhance the intelligence and efficiency of wireless communication networks. In this article, we propose the concept of generative mobile edge networks and overview widely adopted GAI technologies and their applications in mobile edge networks. We then discuss the potential challenges faced by generative mobile edge networks in resource-constrained scenarios. To address these challenges, we develop a universal resource-efficient generative incentive mechanism framework, in which we design resource-efficient methods for network overhead reduction, formulate appropriate incentive mechanisms for the resource allocation problem, and utilize Generative Diffusion Models (GDMs) to find the optimal incentive mechanism solutions. Furthermore, we conduct a case study on resource-constrained mobile edge networks, employing model partition for efficient AI task offloading and proposing a GDM-based Stackelberg model to motivate edge devices to contribute computing resources for mobile edge intelligence. Finally, we propose several open directions that could contribute to the future popularity of generative mobile edge networks.
	\end{abstract}

	\begin{IEEEkeywords}
        6G, generative AI, mobile edge network, Stackelberg game, generative diffusion model.
	\end{IEEEkeywords}
	
	\section{Introduction}
         
    The research and implementation of the 6G technology is significant as it represents the next-generation communication system. In contrast to 5G, 6G networks are completely novel systems that seamlessly integrate air, ground, and satellite communications~\cite{jiang2021road}. This integration offers the potential for broader and global network coverage, decreased latency, increased network capacity, enhanced reliability, and more intelligent network management. 6G networks could provide ultra-reliable and low-latency communication~\cite{chowdhury20206g}, which is crucial for applications that require real-time responsiveness, such as autonomous vehicles, remote surgery, and industrial automation. Moreover, the effective application of 6G technologies possesses significant potential to enable the deployment of large-scale and more complex edge mobile networks.\par

    Mobile edge networks are novel technologies that bring computational capabilities closer to edge devices and end-users, reducing latency and enabling real-time data processing. With emerging technologies such as Artificial Intelligence-Generated Content (AIGC), metaverses, and Internet of Things (IoT), the amount of data in mobile edge networks is substantially upsurging. For example, by the end of 2030, the mobile data traffic of a mobile device will reach 257.1 GB per month, which is 50 times compared to the traffic volume in 2010~\cite{chowdhury20206g}. In future mobile edge networks, features, such as profound intelligence, heightened immersion, and exceptionally low latency, play a pivotal role in greatly enhancing the user experience. Therefore, enhancing the intelligence and efficiency of mobile edge networks holds the potential to sig-\\nificantly empower the future wireless communication system.\par
    
    In recent years, the rapid advancement of computer processing capabilities has spurred the swift development and widespread application of AI across various domains. Notably, as a highly distinctive form of AI, Generative AI (GAI) has gained increasing attention. Unlike conventional AI models, GAI models generate new data instead of discriminating existing data in the majority of instances. Cutting-edge GAI models, for example, the recently acclaimed ChatGPT\footnote{\url{https://chat.openai.com/}}, which is a transformer-based Large Language Models (LLMs), possess remarkable generalization capabilities and creative prowess. GAI excels at producing a wide range of content and data, including images, text, videos, and more, in response to user inputs and prompts. For instance, in vehicular metaverses, GAI can be utilized to generate real-time road tracks and virtual pedestrians for AR navigation tasks~\cite{kang2023adversarial}.\par
    
     Currently, mainstream mobile AIGC services are predominantly executed in the cloud, generating content and transmitting it to users. However, cloud-based AIGC service can introduce significant service latency, which may be unacceptable for certain mobile AI generation tasks that demand real-time immersion and exceptionally low latency. The integration of GAI with mobile edge networks is giving rise to a novel paradigm named \textit{generative mobile edge networks}. Generative mobile edge networks are a novel type of edge network that possess powerful data generation and decision-making capabilities, which hold immense potential for improving the intelligence of networks and offering substantial benefits. For instance, deploying GAI models on mobile edge devices can alleviate the computational overhead on the cloud. In addition, GAI models, such as Generative Diffusion Models (GDMs), have shown significant potential in the realm of network optimization~\cite{du2023beyond}. This is attributed to their capability to capture intricate and high-dimensional network environment distributions, thereby facilitating more sophisticated and efficient decision-making processes. Although generative mobile edge networks hold significant potential in revolutionizing the future of mobile AIGC services and applications, there exist several challenges in enabling generative mobile edge networks~\cite{xu2023unleashing}, including the substantial resource overhead associated with deploying GAI models on edge devices, constraints on bandwidth resources when requesting AIGC services, lower model accuracy with resource constraints, and dynamic network environments affecting network optimization and decision-making. \textit{To the best of our knowledge, this is the first work that presents the concept of generative mobile edge networks and provides innovative research for enabling generative mobile edge networks.} Our contributions can be summarized as follows:
     \begin{itemize}
	\item  We provide a comprehensive analysis of the GAI technologies and their applications in the generative mobile edge networks. Then, we discuss the main challenges in enabling generative mobile edge networks.
	\item  We present a novel resource-efficient generative incentive mechanism framework for generative mobile edge networks, in which we design resource-efficient methods for network overhead reduction and formulate GDMs-based incentive mechanisms.
	\item  We conduct a case study on resource-constrained edge intelligence scenarios, in which we design a Stackelberg game mechanism to effectively allocate network resources. Furthermore, we adopt GDMs to generate the optimal game solution. Numerical results demonstrate that the proposed GDM-based method outperforms other baseline.\textit{ To the best of our knowledge, we are the first to apply GDMs in the Stackelberg game. }
	\end{itemize}

 \begin{figure*}[t]
    \centerline{\includegraphics[width=0.9\textwidth]{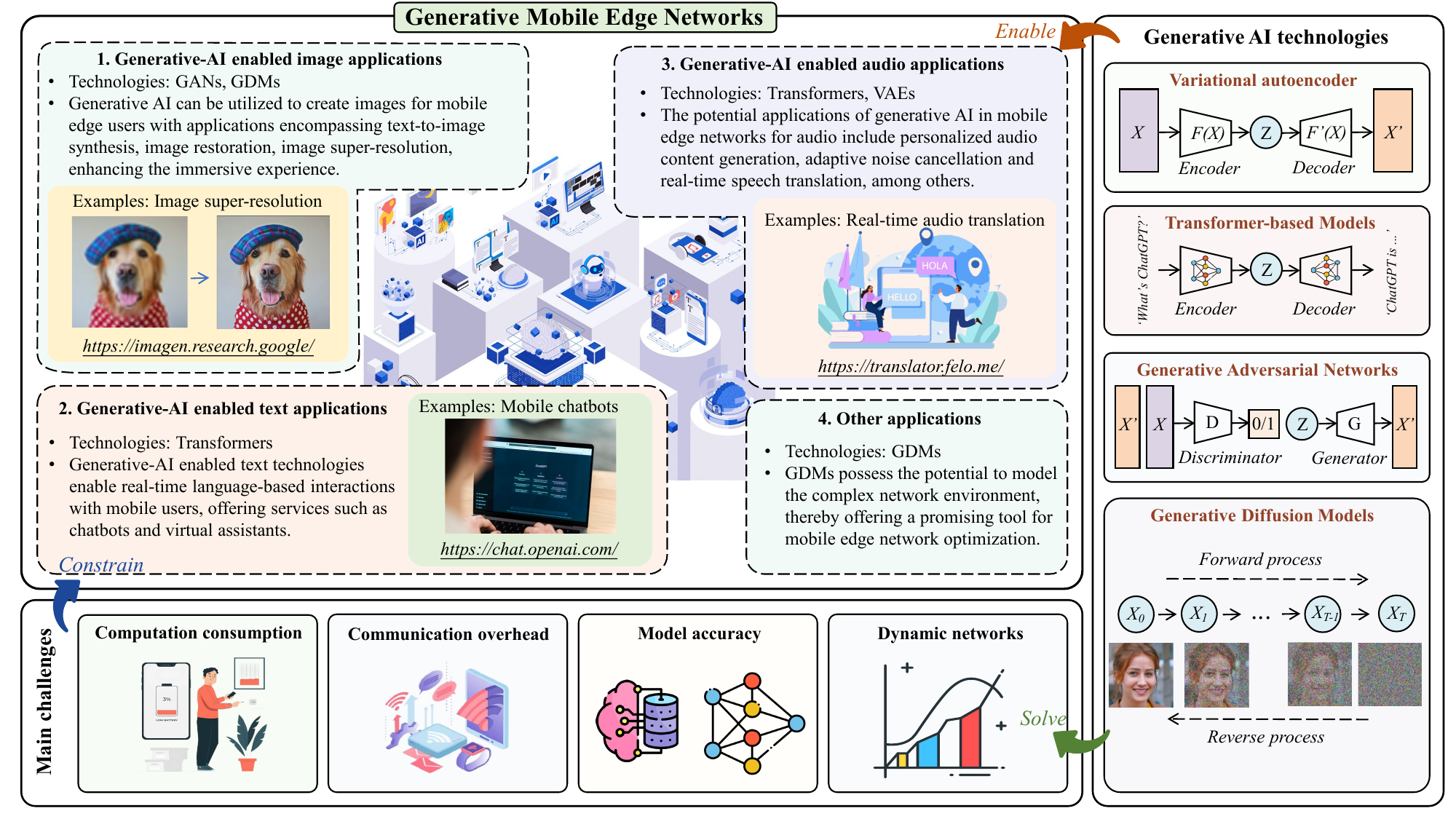}}
    \captionsetup{font=footnotesize}
    \caption{The schematic of generative mobile edge networks. We summarize four potential GAI-enabled applications and provide examples. Additionally, we discuss several GAI techniques that enable generative mobile edge network applications and identify the main challenges that constrain the development and widespread adoption of such applications.}
    \label{fig1}
\end{figure*}
        
 \section{Generative Mobile Edge Networks}
In this section, we introduce several main GAI techniques in mobile edge networks. Subsequently, we present various GAI applications in mobile edge networks. Finally, we provide a discussion of the main challenges and considerations that arise when implementing GAI in resource-constrained mobile intelligent networks. Figure \ref{fig1} illustrates the schematic of generative mobile edge networks.
 \subsection{Fundamental Generative AI Techniques}
We review several fundamental generative AI techniques, including their architectures and applications in mobile edge network scenarios.
 \begin{itemize}
	    \item  \textbf{Generative Adversarial Networks (GANs):}
            As a fundamental model of GAI technologies, GANs consist primarily of a generative network and a discriminative network. The learning process of GANs involves adversarial training between these two networks~\cite{karras2018progressive}. While GANs possess excellent data generation ability, the unstable nature of GANs leads to slow convergence and occasional failure to converge. GANs can be utilized for data augmentation in mobile edge networks, which enhance the data processing performance in edge networks~\cite{lin2023unified}.
		\item  \textbf{Generative Diffusion Models (GDMs):}
            GDMs are trained to learn the reverse diffusion process to generate the desired data by denoising the Gaussian noise~\cite{ho2020denoising}. GDMs not only show great potential in image generation but also hold significant prospects in network optimization~\cite{du2023beyond}. In particular, GDMs are capable of finding the optimal solution in high-dimensional and complex network environments effectively.
		\item  \textbf{Variational Autoencoders (VAEs):}
            VAEs are a typical generative network with an encoder-decoder structure. The encoder maps the input data into a latent space, while the decoder reconstructs new data samples from this latent space, capturing a distribution that closely resembles the original data~\cite{harshvardhan2020comprehensive}. VAEs can be used for data compression in mobile edge networks. By encoding data into a lower-dimensional latent space, VAEs can reduce the size of data before transmission, minimizing bandwidth usage and improving network efficiency.
            \item  \textbf{Transformers:}
            Transformers, being a crucial foundation model for natural language generation, have significantly advanced the development of LLMs like ChatGPT. By leveraging self-attention mechanisms, Transformers effectively capture the relations among elements of sequential data, making them highly suitable for various natural language processing tasks~\cite{xu2023unleashing}. In generative mobile edge networks, transformers possess the potential to enable real-time language-based interactions with mobile users, offering services such as chatbots and virtual assistants.
	\end{itemize}
 \subsection{Generative AI-enabled Applications in Generative Mobile Edge Networks}
We systematically provide several GAI-enabled applications in generative mobile networks, which can demonstrate the potential power of generative mobile edge networks and serve as potential future research directions in enabling generative mobile edge networks.
 \begin{itemize}
	    \item  \textbf{GAI-enabled text applications:}
            GAI-enabled text applications enable real-time language-based interactions with mobile users, offering intelligent and convenient GAI services such as chatbots and virtual assistants, language translation, and creative writing assistance~\cite{du2023enabling}. For example, LLMs like ChatGPT, trained on extensive text data, exhibit the ability to generate human-like text responses based on given prompts. In generative mobile edge networks, the application of AI-generated text technologies can provide mobile users with real-time and human-like content. For instance, it can offer real-time driving navigation in Internet of Vehicles (IoV) scenarios.
		\item  \textbf{GAI-enabled image applications:}
            GAI-generated image applications hold great significance in generative mobile edge networks. Specifically, GAI has the capability to generate novel images based on user prompts, with applications encompassing text-to-image synthesis, image restoration, image super-resolution, and more~\cite{xu2023unleashing}. An example of the GAI-enabled image application is the generation of road trajectory prediction images for mobile users in the IoV domain~\cite{du2023enabling}. Additionally, GAI can further enhance the clarity of blurry images and elevate user immersion by leveraging super-resolution techniques\footnote{\url{https://imagen.research.google/}}, particularly when utilizing Internet of Things technologies to acquire real-time image information of road conditions ahead.
            \item  \textbf{GAI-enabled audio applications:}
            GAI widely adopted in audio generation possesses the capability to tailor audio data generation based on user requirements, which is particularly significant and valuable in mobile edge networks. For instance, in GAI-enabled audio applications, GAI can provide personalized voice assistant services to users~\cite{wen2023generative}. In addition, GAI can also customize music for individual users, thereby enhancing their listening experiences. Another promising GAI-enabled audio application in mobile edge networks is real-time audio translations\footnote{\url{https://translator.felo.me/}}, which has the potential to significantly enhance the interactive experience between edge users and their environment.
            \item  \textbf{Other applications:}
            The applications of GAI in generative mobile edge networks extend beyond the mentioned areas. There are various other research directions that can leverage the capabilities of GAI. One such direction is network optimization, where optimization tools based on GDMs can effectively model the state and action spaces of complex environments, providing a promising approach for novel optimization techniques.
	\end{itemize}
\subsection{Main Challenges in Enabling Generative Mobile Edge Networks}
Mobile Edge intelligence plays a pivotal role in the effective implementation of generative mobile edge networks. Specifically, the deployment of generative AI on mobile edge networks often relies on techniques such as distributed machine learning and edge inference~\cite{xu2023unleashing}, which serve as integral components of edge intelligence. Therefore, from the perspective of network resource constraints, we provide several main challenges in enabling generative mobile edge networks.
 \begin{itemize}
 		\item  \textbf{Huge computation consumption for model training and inference:}
            In conventional edge intelligence scenarios, the constrained power of edge devices may pose challenges in affording the substantial computational resource consumption required for AI training and inference. Furthermore, the performance of edge networks is often impacted by the heterogeneity of computing resources on edge devices. In generative mobile edge networks, GAI models tend to be larger in scale compared to traditional discriminative models like convolutional neural networks. Additionally, due to the instability of generative models and the distributed characteristic of mobile edge networks, AI models may need to be trained multiple times to generate appropriate results that meet user needs~\cite{du2023beyond}, leading to generative mobile edge networks consuming computational resources more than before.
		\item  \textbf{High communication overhead for model training and inference:}
            In generative mobile edge networks, the training and inference process of GAI models may involve model transmission. For instance, in the training process of Federated Learning (FL), edge devices would transmit their local models to the server for global model aggregation during each round of global iteration. This process typically necessitates several hundred epochs to achieve model convergence, causing excessive communication delay in model training. Considering the heterogeneity of devices and the variability of wireless communication network conditions, it is essential to reasonably allocate communication resources like bandwidth to edge devices. Moreover, when mobile users request AIGC services from the server, it is imperative to thoroughly study the trade-off problem between the quality of AIGC services and communication efficiency~\cite{du2023enabling}.
            \item  \textbf{Low model training and inference accuracy:}
            Given limited network resources, how to improve the accuracy of the model as much as possible while maximizing the utility of resources is a highly significant and challenging problem~\cite{xu2023unleashing}. Generative mobile edge networks, in particular, face even more intricate issues compared to traditional edge networks. This complexity is primarily driven by the inherent instability of GAI models. The growing creative capability of GAI models has greatly enriched personalized and customized AIGC services, which has simultaneously added complexity to the evaluation of model accuracy, rendering network optimization a challenging endeavor. 
            \item  \textbf{Dynamic environments of generative mobile edge networks:}
             The arrival of 6G will bring significant growth in the scale of mobile edge networks, leading to higher dimensionality and complexity. Therefore, how to effectively allocate resources in a heterogeneous and dynamically changing network poses a serious challenge. Traditional optimization principles and tools often rely on accurate and comprehensive network information. However, it is difficult to obtain such information in a complex network environment~\cite{wen2023generative}. 
             Although the excellent decision-making ability of Deep Reinforcement Learning (DRL) has shown promising prospects in complex network optimization, DRL also has certain limitations. One such limitation is that when the state space or action space becomes too large, DRL may not be able to find the optimal network solution effectively~\cite{du2023beyond}.
	\end{itemize}
 
 \begin{figure*}[ht]
    \centering  
    \centerline{\includegraphics[width=0.9\textwidth]{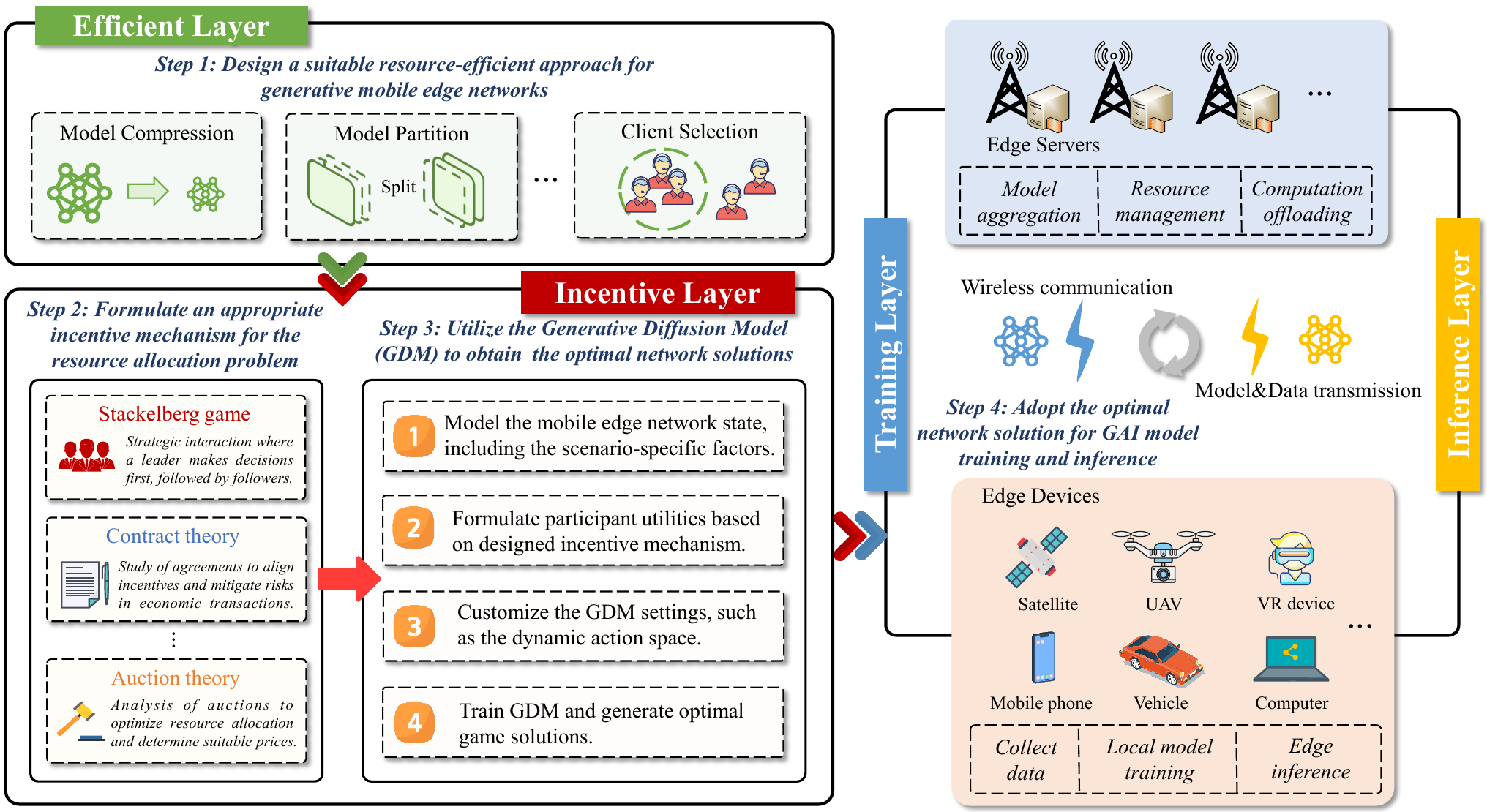}}
    \captionsetup{font=footnotesize}
    \caption{Resource-efficient Generative AI-based incentive mechanism framework for generative mobile edge networks. The proposed framework consists of three layers, i.e., an efficient layer, an incentive layer, and a training$\&$inference layer.}
    \label{fig2}
\end{figure*}
\section{Resource-efficient Generative Incentive Mechanism Framework}
In prior studies, numerous resource-efficient methods have been proposed to tackle the resource constraint problem of edge intelligence and enhance the performance of traditional mobile networks~\cite{yu2021toward}. However, for generative mobile edge networks, further exploration is necessary. In this section, we commence by presenting resource-efficient methods that are widely applied in various edge intelligence scenarios. Subsequently, we introduce several incentive mechanisms that are widely employed to effectively address resource allocation problems. Lastly, we propose a general resource-efficient generative incentive mechanism framework to address the aforementioned challenges.
\subsection{Resource-efficient Methods in Mobile Edge Intelligence Networks}
\begin{itemize}
	    \item  \textbf{Model compression:}
            With the increasing size and complexity of AI models, a range of model compression techniques, such as model pruning and model quantization, have been adopted for edge networks. These techniques are designed to enhance the efficiency of network resource utilization without substantial compromise on model performance. Specifically, model Pruning reduces the number of model parameters by eliminating redundant parameters, while model quantization achieves lightweight models by decreasing the precision of model parameters~\cite{xu2023unleashing}. Given the complexity and instability of GAI models, there is an urgent need for more efficient and customized compression techniques to enable generative mobile edge networks.
		\item  \textbf{Model partition:}
            Model partition is another widely adopted method to alleviate the resource pressure on edge intelligent networks, which typically involves splitting a deep learning model into two parts~\cite{zhou2019edge}. During the training or inference phases of edge AI, these splitting models are deployed separately on the edge server. When the server is burdened with too many computing tasks, AI tasks can be offloaded to the edge devices if necessary to alleviate the computational load on the server. If the cost of executing AI tasks on edge devices becomes too high, the server can be requested to assist with another part of the task, which is also applicable to generative mobile edge networks. Especially, the authors in ~\cite{xu2023unleashing} proposed a method for the partition of diffusion models.
		\item  \textbf{Client selection:}
            Distributed machine learning approaches, such as FL, involve collaborative training with a large number of edge clients during the model training process. However, in practice, attempting to consider all devices in the model training may yield unsatisfactory results~\cite{zhou2019edge}, which leads to convergence slowdown or even degradation, while simultaneously increasing the resource consumption of the network. Client selection is influenced by various factors, including but not limited to resource heterogeneity, dynamic communication environments, and the quality of local data. To enable generative mobile edge networks, it is essential to devise appropriate resource allocation strategies for client selection.
             
	\end{itemize}
\subsection{Incentive Mechanism Designs in Mobile Edge Networks}
Designing suitable incentive mechanisms as resource allocation schemes holds the potential to significantly enhance the performance of generative mobile edge networks, which can increase the overall profitability of the network from the economic perspective~\cite{du2023beyond}. Efficient incentive mechanisms can also incentivize more users to engage with the network, leading to greater profits and a more balanced ecosystem. In the following part, we introduce several incentive mechanisms that are widely utilized in mobile edge networks.

\begin{itemize}
	    \item  \textbf{Stackelberg game:}
            In a mobile edge network, users may compete for access to limited network resources. Stackelberg games can facilitate decision-making for users regarding how much network bandwidth or computing resources they are willing to share, considering their self-interest and the potential for cooperation~\cite{zhang2023learning}. The interaction between leaders and followers in decision-making will iterate until reaching the Nash equilibrium.
		\item  \textbf{Contract theory:}
            Contract theory, as a highly effective incentive mechanism, is extensively employed in mobile networks. For instance, in the generative mobile edge network, an AIGC service provider assumes the role of an employer and makes contracts with AIGC users, who function as employees. Optimal contracts designed based on contract theory can proficiently address issues arising from information asymmetry~\cite{du2023beyond}, thereby mitigating unfair trading resulting from the self-serving strategies of employees.
		\item  \textbf{Auction:}
            Auction theory can effectively incentivize competition in mobile edge networks under limited network resources. For example, the AIGC service provider acts as an auctioneer, setting initial prices for mobile users to bid on AIGC computing resources, and selecting successful auction winners based on their bid prices and requirements. In summary, auction theory offers an efficient resource allocation and pricing mechanism for mobile networks~\cite{wen2023generative}.
	\end{itemize}

\subsection{Framework Designs}
As shown in Fig. \ref{fig2}, we propose a general resource-efficient generative incentive mechanism framework for enabling generative mobile edge networks. The framework comprises three layers, i.e., a resource-efficient layer, an incentive mechanism layer, and a model training \& inference layer. The detail of the framework is provided as follows:
\begin{itemize}
	    \item  \textit{\textbf{Step 1. Design a suitable resource-efficient approach for mobile edge intelligence:}}
        In the resource-efficient layer, the servers, as network operators, focus on addressing resource constraints in mobile edge intelligence. Here, tailored resource-efficient edge intelligence methods based on individual user requirements and resource constraints will be designed, e.g., customized model compression methods for FL scenarios. It is important to emphasize that each of the different resource-efficient approaches involves variables that need to be optimized, such as the compression rate in model compression or the splitting points in model partition.
            \item  \textit{\textbf{Step 2. Formulate an appropriate incentive mechanism for optimal resource allocation:}}
            Based on the designed resource-efficient approach, it is crucial to further efficiently and reasonably allocate resources throughout the entire network to maximize overall benefits. This involves formulating the most appropriate incentive mechanism tailored to the current network environment and user requirements. Throughout the design process, various network factors must be taken into account, such as compression rates, model accuracy, resource constraints, and more.
            \item  \textit{\textbf{Step 3. Utilize the GDM to obtain the optimal network solution:}}
            To tackle the well-designed incentive mechanism, it's essential to identify the optimal strategy for problem optimization. While previous optimization schemes traditionally rely on methods like convex optimization and real-world network environments tend to be intricate and dynamic. Consequently, GDMs as novel optimization algorithms can be utilized to effectively simulate complex real-world scenarios, thereby obtaining more stable and efficient solutions. 
            \item  \textit{\textbf{Step 4. Adopt the optimal network solutions for model training and inference:}}
            After deriving the optimal incentive mechanism strategy by utilizing the GDM, we can implement this approach in edge intelligence scenarios, such as distributed machine learning for model training and inference. By applying our proposed framework, it can effectively address resource scheduling challenges in resource-constrained mobile edge networks, thereby enhancing the overall network efficiency and performance.
	\end{itemize}

\begin{figure*}[ht]
    \centerline{\includegraphics[width=0.9\textwidth]{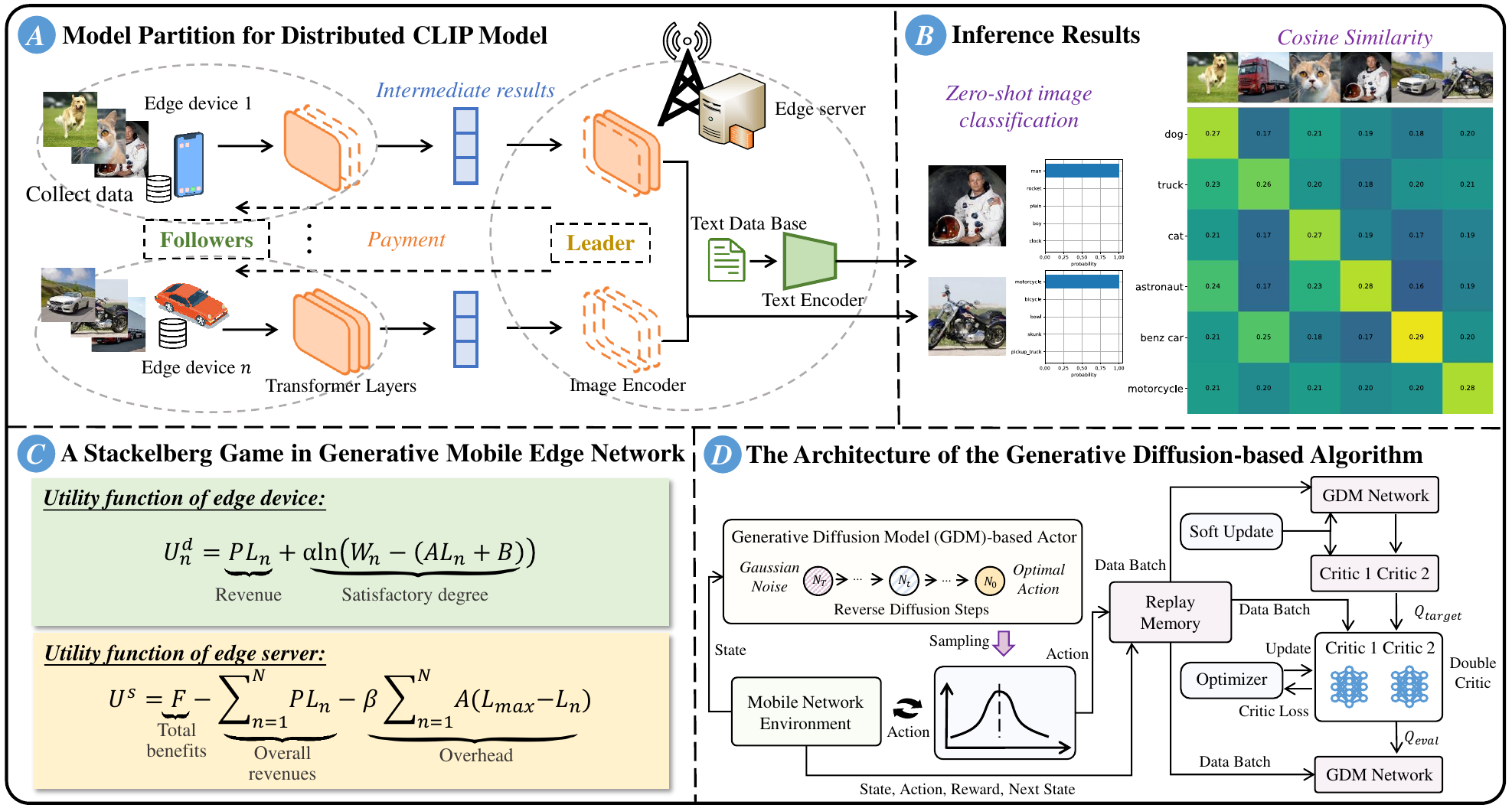}}
    \captionsetup{font=footnotesize}
    \caption{A case study on generative diffusion-based Stackelberg game in mobile edge intelligence. Part A represents the model partition approach for the distributed CLIP model in mobile edge networks. Part B illustrates the inference results of the CLIP model, which completes the zero-shot image classification by calculating the cosine similarity between the image tensor and text tensor. Part C shows the Stackelberg game in the distributed CLIP model. Part D demonstrates the architecture of the GDM-based algorithm, which is adopted to identify the optimal game solution.
    }
    \label{fig3}
\end{figure*}
 
\section{Case Study: Generative Diffusion-based Stackelberg Game in Mobile Edge Intelligence}
In this section, we conduct a case study on mobile edge intelligence. Specifically, we propose a GDMs-based Stackelberg game model, which can fairly incentivize users to contribute their computing resources for mobile edge networks. 
\subsection{System Model}
In conventional centralized mobile networks, edge devices typically accumulate substantial volumes of raw data, which is subsequently uploaded to the server for performing AI tasks like image recognition and object detection. However, due to the limitations of edge server resources, simultaneously processing numerous AI tasks with extensive data can lead to considerable overhead. To address this challenge and offload AI tasks from edge servers to edge devices, we propose the integration of model partition and the Contrastive Language–Image Pre-training (CLIP) model\footnote{\url{https://openai.com/research/clip}} named distributed CLIP model. The CLIP model is an open-source model developed by OpenAI, capable of tasks like zero-shot image classification without the need for fine-tuning. For the distributed CLIP model, we split the transformer-based image encoder of the CLIP model into two parts, with the first part residing on the edge device and the second part residing on the edge server. The image data collected by the edge device are first encoded by the mobile-side encoder. Then, the edge device sends the intermediate results to the server. the intermediate results are encoded by the other half of the image encoder to generate the image vector. Finally, this vector is compared to the encoded text vector from the text knowledge base to determine the probability of the image category.

\subsection{Generative Diffusion-based Stackelberg Game Solutions}
We consider that each edge server can support $N$ edge devices. Each edge device is equipped with a transformer-based image encoder, which comprises a total of $L_{max}$ stacked transformer layers.
 When conducting edge inference, each edge device chooses a specific number of layers, denoted as $L_n$, for local inference. After that, the edge device transmits the encoded intermediate results to the edge server, then the remaining process will be completed by the server. \par
\textit{1) Edge device utility:} 
The utility of each edge device is represented by $U^d_n$, which comprises two parts. The first part is the revenue generated by offering computing services, where $P$ denotes the pricing strategy for incentivizing edge devices. The second part is a satisfactory degree. Taking into account the self-interested nature of edge devices, we consider that the satisfaction degree of an edge device increases monotonically as it allocates more computation for itself after completing the edge computing task. As shown in Fig.\ref{fig3}, $W_n$ represents the maximum amount of computation device $n$ can accommodate. The relationship between the amount of computation and the number of layers can be approximated as a linear relationship~\cite{han2021transformer}, therefore the amount of computation can be expressed as $(AL_n+B)$, where $A>0$ and $B>0$ are the corresponding parameter. Note that $\alpha$ denotes the tradeoff parameter between the satisfactory degree and monetary rewards.
\par
\textit{2) Edge server utility:} 
For the edge server, its utility $U^s$ is equal to the difference between the total revenue derived from conducting edge AI inference and its incurred overhead. In this context, we consider that the total revenue remains fixed and is represented by $F$. The overhead encompasses the overall compensation paid to each edge device and the expenses associated with completing any remaining computing tasks. Note that $\beta$ represents the economic coefficient related to the edge server's computing tasks.
\par
\textit{3) Stackelberg game formulation:} 
In this scenario, edge devices and edge server seek to maximize their individual profits, and their interaction can be considered as a monopoly market. Precisely, edge server has pricing authority, while edge devices respond to the price and decide the extent of computing resources they contribute to the network based on their specific needs. When the price decided by the edge server is too low, edge devices might be disinclined to contribute their computing resources, thus failing to alleviate the computational burden of the edge server. Conversely, as the price increases, edge devices become more motivated to contribute computing resources in pursuit of higher profits for themselves. Motivated by the above analysis, we propose a Stackelberg model to maximize the benefit of the edge server and maintain its monopoly power~\cite{zhang2023learning}, where the Stackelberg game has been widely utilized by the monopolist to strategically set the price. The Stackelberg game between the edge server and edge devices consists of two stages. In the first stage, the edge server, acting as the leader, sets the price for each transformer layer to maximize the overall utility. In the second stage, edge devices as followers determine the number of transformer layers they employ for inference to maximize their utilities.
\par
\subsection{GDM-based Stackelberg Game Solution}
The process of utilizing the GDMs to find the optimal Stackelberg game solution in generative mobile edge networks can be summarized as follows:
\begin{itemize}
    \item  \textbf{Step 1. Model the network environment:}
    In the Stackelberg game, the environment of the mobile edge network is initially defined. The current environmental state consists of the pricing of the edge server and the number of transformer layers for edge devices. In the current round $j$, the state can be represented as $S^j = \{p^j, L^j_1,\ldots, L^j_N \}$. It is worth mentioning that the initial state space can be generated through random sampling.
    \item  \textbf{Step 2. Formulate the utilities of participants:}
    In the second step, we define the utility functions of the edge server and edge devices, respectively. The utility function of edge devices is used to calculate the optimal strategy for edge devices. Once the optimal strategy for edge devices is determined, the edge server can calculate its total utility $U^s$, which serves as the objective function that the edge server aims to maximize.
    \item  \textbf{Step 3. Customize the GDM settings:}
    The third step is to customize the GDM settings based on the proposed Stackelberg game. The action space of the GDM encompasses all possible price strategies for the edge server. Regarding the reward function, a common approach is to assign a reward based on the change in the utility of the edge server between the current state and the previous state. In the current round, when the utility is greater than the utility of the previous state, the reward can be set to 1. Conversely, if the current utility is lower, the reward can be set to 0. This setup incentivizes the GDM to learn strategies that increase utility over time.
    \item  \textbf{Step 4. Train the GDM and generate optimal game solution:} 
    After completing the training of the GDM for the Stakelberg game, it becomes capable of finding the optimal strategy based on any given state. Unlike traditional DRL, GDMs employ a different approach to generate optimal solutions. Instead of utilizing back-propagation algorithms to optimize model parameters directly, GDMs aim to generate the optimal Stackelberg game solution by denoising the initial distribution.

\end{itemize}

\subsection{Numerical Results}
\begin{figure}[h]
    \centering  
    \includegraphics[width=0.4\textwidth]{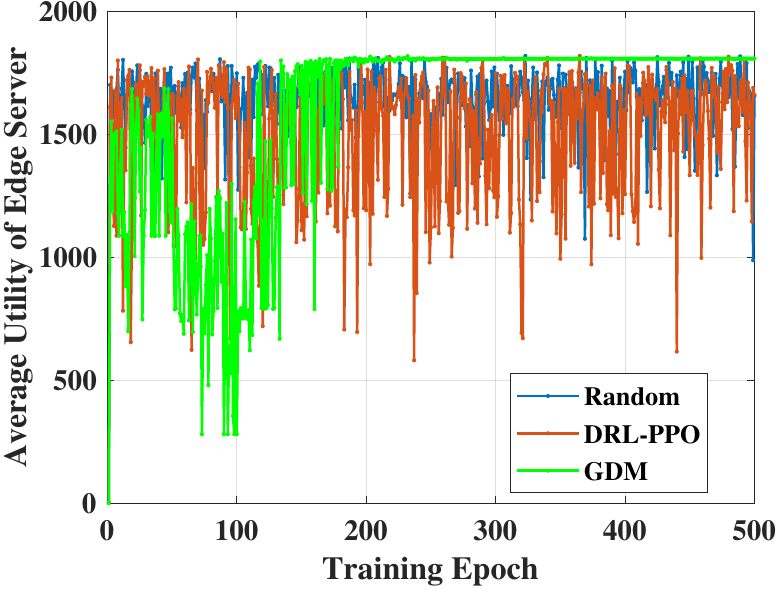}   
    \captionsetup{font=footnotesize}
    \caption{Average utility values versus the number of training epochs. For performance comparison, we show the results of GDM, random, and DRL-PPO algorithms.}
    \label{fig4}
\end{figure}

\begin{figure}[h]
    \centering  
    \includegraphics[width=0.4\textwidth]{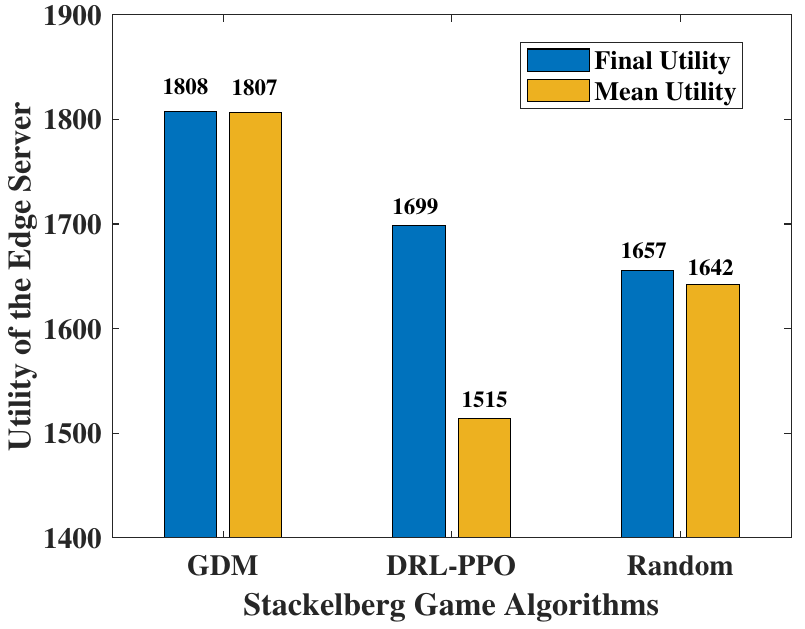}   
    \captionsetup{font=footnotesize}
    \caption{Comparison of the final utility and the mean utility with the different methods.}
    \label{fig5}
\end{figure}

Figure \ref{fig4} shows the utility curves of three types of game solution-finding strategies. To fairly evaluate the performance, we conducted the experiments three times and took the average utility of the edge server. While both DRL-PPO and random methods exhibit larger fluctuations, GDMs achieve higher utility and better convergence under the same parameter settings. This is attributed to the fact that in scenarios with incomplete network information, GDMs are capable of simulating the data distribution of dynamic and complex network environments more effectively~\cite{du2023beyond}, resulting in better performance and stability.
Figure \ref{fig5} further illustrates that the proposed GDM-based game solution-finding scheme outperforms baseline methods in terms of both the final utility and mean utility values of the edge server. For the mean utility, we consider a range from 200 to 500 epochs. In general, the numerical results show that the proposed GDM-based game solution generation scheme outperforms the baseline methods.

\section{Future Directions}
\subsection{Generative AI for Semantic Communications}
In the context of edge intelligence, semantic communication is a promising new communication paradigm. The powerful generation ability of GAI can be used to enhance the performance and efficiency of semantic communication. GAI, such as GANs, has been employed to enhance data for AI-based semantic communication~\cite{lin2023unified}, aiming to minimize the retraining expenses of semantic communication systems in out-of-distribution situations. The design of improved semantic communication systems plays a vital role in reducing delays in mobile AIGC services.
\subsection{Secure AIGC Services in Generative Mobile Edge Networks}
While GAI offers numerous advantages, it also encounters certain security challenges. Specifically, GAI can be exploited to generate deceptive and misleading content, and the misuse of deepfakes and similar technologies can adversely affect the AIGC service experience of mobile users. Consequently, ensuring the security of mobile users when utilizing AIGC services poses a significant challenge. Thus, future research on GAI in mobile networks can focus on developing robust defense mechanisms and identifying fraudulent content to address this issue effectively.
\subsection{Privacy Preservation in Generative Mobile Edge Networks}
 Privacy leakage is also a critical challenge faced by generative mobile edge networks. When requesting cloud-based AIGC services, mobile users need to send some prompts to the edge server, which may contain user privacy information, thus increasing the risk of privacy leakage. Additionally, during the pre-training of GAI, many large models acquire training data through web crawlers, which may inadvertently include private information. Consequently, future research in enabling generative mobile edge networks can focus on developing methods to provide AIGC services while safeguarding the privacy of mobile users. Moreover, efforts should be intensified to enhance the security evaluation of AIGC large model services, thereby preventing privacy leakage effectively.
\bibliographystyle{ieeetr}

\bibliography{ref}

\end{document}